\documentclass[aps,prl,twocolumn,groupedaddress,showpacs]{revtex4-1}
\usepackage{amsmath,amssymb}
\usepackage{graphics,graphicx}
\usepackage{color}

\newcommand{\pd}[2]{\frac{\partial#1}{\partial#2}}

\newcommand{\braket}[1]{\langle#1\rangle}
\newcommand{\der}[2]{\frac{d#1}{d#2}}

\begin{document}

%\preprint{}

\title{Upconversion of optical signals with multi-longitudinal-mode pump lasers}


\author{J.~S.~Pelc$^{1}$, G.-L.~Shentu$^2$, Q.~Zhang$^{2,\ast}$, M.~M.~Fejer$^1$, and Jian-Wei Pan$^2$}
\affiliation{$^1$ E.~L.~Ginzton Laboratory, Stanford University, 348 Via Pueblo Mall, Stanford CA 94305, USA\\
$^2$ Shanghai Branch, National Laboratory for Physical Sciences at Microscale and Department of Modern Physics, University of Science and Technology of China, Shanghai, 201315, China\\
$^\ast$ Corresponding author: qiangzh@ustc.edu.cn}
\date{\today}


\date{\today}

\begin{abstract}
Multi-longitudinal-mode lasers have been believed to be good candidates as pump sources for optical frequency conversion.  However, we present a semi-classical model for frequency conversion of optical signals with a multimode pump laser, which shows that fluctuations of the instantaneous pump power limit the conversion efficiency.  In an experiment, we upconverted a 1.55-$\mu$m optical signal in a periodically poled lithium niobate waveguide using with a multi-longitudinal-mode laser, an observed a maximum conversion efficiency of 70\%, in good agreement with our theoretical model.  Compared to single-mode pumping, multimode pumping is not a suitable technique for attaining stable near-unity-efficiency frequency conversion.  However, the results obtained here could find application in characterization of the spectral or temporal structure of multi-longitudinal-mode lasers.
\end{abstract}


\maketitle

\section{Introduction}

Dating to the earliest explorations of nonlinear optics, upconversion has been proposed as a method by which one may achieve high-sensitivity detection of weak optical signals in the infrared spectral region \cite{armstrong_interactions_1962,midwinter_up_conversion_1967}.  Although it has been used for upconversion of images in astronomy, the low quantum efficiencies (in the range of fractions of a percent, limited by the available pump lasers and nonlinear media) precluded widespread use \cite{midwinter_up_conversion_1967,boyd_infrared_1977}.  With the development of periodically poled lithium niobate (PPLN) waveguides, conversion efficiencies exceeding 90\% (limited by waveguide propagation losses) have been attained \cite{roussev_periodically_2004}.  One technologically important application of upconversion in recent years has been for single-photon detection.  Upconversion has been shown to preserve the quantum properties of a light field, and therefore can enable the noiseless frequency conversion of single-photon Fock states \cite{kumar_quantum_1990}.  Because single-photon detectors for the visible and near-infrared spectral range based on Si avalanche photodiodes (APDs) have many advantageous properties over those consisting of InGaAs/InP for the 1.55-$\mu$m telecommunications band, upconversion is a promising technique by which the spectral range of Si APDs may be extended to wavelengths longer than the Si bandgap \cite{hadfield_single-photon_2009}.  In an upconversion single-photon detector, an incident signal in the infrared at $\omega_1$ is combined in a $\chi^{(2)}$ medium with a strong pump at $\omega_p$ to produce radiation at $\omega_2 = \omega_1 + \omega_p$ via the process of sum-frequency generation (SFG). Recently, upconversion single photon detection has found many important applications in quantum cryptography, near infrared spectroscopy, and single-photon manipulation \cite{Zhang_MegabitsQKD_2009,zhang_waveguide-based_2008,rakher_quantum_2010}.

Nearly all upconversion experiments have used a single mode laser as a pump. In comparison, multimode lasers could be cheaper and provide higher power. In addition, laser technology in some spectral windows, for example 1.8-$\mu$m band, is not as well developed as in the NIR around 1~$\mu$m.  While high-power multi-longitudinal-mode lasers are available in this spectral range, high-power economical single-mode sources do not yet exist. However, a pump source in this spectral region is needed for low-noise 1.55-$\mu$m band upconversion single photon detection because both non-phasematched parametric fluorescence \cite{pelc_influence_2010} and spontaneous Raman scattering \cite{pelc_long-wavelength-pumped_2011} are problematic for pump frequencies $\omega_p > \omega_1$.

A recent paper has demonstrated the upconversion of single-photon-level signals with a multi-longitudinal-mode pump laser at 1.06~$\mu$m \cite{pan_single-photon_2008}.  The authors reported a conversion efficiency of 96\% using a laser in which approximately 400 axial modes were lasing.

In this paper, we describe a semiclassical model for the upconversion of optical signals, and find that  multi-longitudinal-mode pumping is in general not suitable for applications requiring stable near-unity conversion efficiency. The instantaneous power output of a multi-longitudinal-mode laser fluctuates in time as the axial modes beat with one another.  If these fluctuations are slower than the time scale of the upconversion process (defined by the group-velocity walkoff between the interacting waves), then the conversion efficiency will also be a rapidly fluctuating function of time.  We report on experiments on upconversion in PPLN waveguides using a multi-longitudinal-mode Tm-doped fiber laser at 1.94~$\mu$m.  The conversion efficiency is consistent with the mode-beating model we develop, and additional measurements show that the frequency content of the upconversion pump source is transferred onto the depleted signal.  Our theory and observations, contrasting with an earlier presentation in \cite{pan_single-photon_2008}, show that multi-longitudinal-mode pumping is in general not a suitable technique for frequency conversion of optical signals. However, our results do show that upconversion can be an effective way to characterize the spectral or temporal structure of a laser source in a spectral region where sufficiently fast detectors are unavailable.

\begin{figure*}[t]
\includegraphics[width=0.98\textwidth]{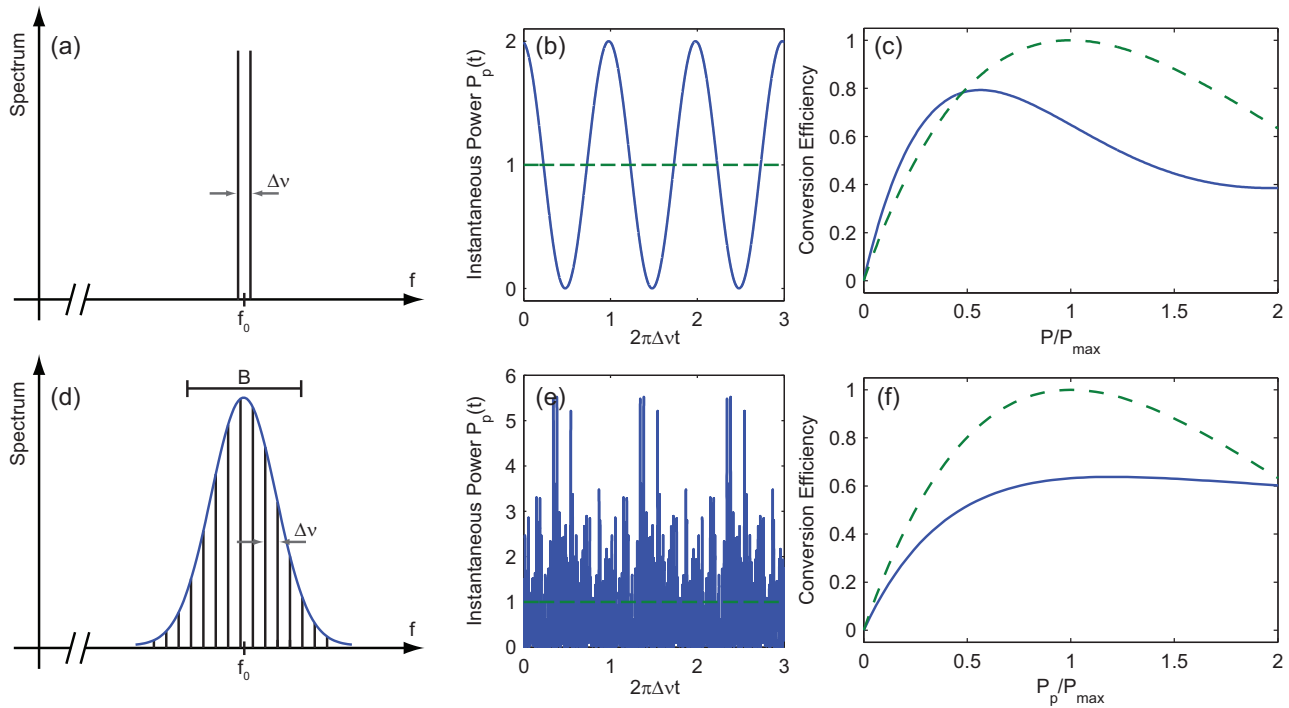}%
\caption{\label{fig:theory}(a) Spectral and (b) temporal representation of a two-mode laser (solid) with fixed phases between the oscillating modes; (c) conversion efficiency computed with Eq.~(\ref{eq:conv-eff}) versus normalized average pump power $\bar{P}_p/P_\mathrm{max}$ for two-mode (solid) and single-mode (dashed) pump lasers. (d-f) Same as (a-c) for a pump laser with the spectral characteristics used in the experiment.}
\end{figure*}

\section{Theory}

We seek to describe the upconversion of an optical signal $a_1(t)$ at center frequency $\omega_1$ via the interaction with a strong optical pump at center frequency $\omega_p$, to the target frequency $\omega_2 = \omega_1 + \omega_p$, in a periodically poled nonlinear waveguide of length $L$.  We consider plane-wave (or single-spatial-mode waveguide) interactions, in the limit of low optical losses and negligible pump depletion.  The coupled wave equations for the signal and target fields are written as
\begin{align} \label{eq:svea}
\pd{a_1}{z} + \delta \nu_{1p} \pd{a_1}{\tau} &= -i \kappa a_p^\ast a_2 e^{-i \Delta k z}\nonumber\\
\pd{a_2}{z} + \delta \nu_{2p} \pd{a_2}{\tau} &= -i \kappa a_p a_1 e^{i \Delta k z}
\end{align}
where the field amplitudes are normalized as $[|a_j(z,t)|^2] = $ photons s$^{-1}$, and the normalized time variable $\tau = t - z/u_p$ (i.e. a frame co-propagating with the pump, where $u_p$ is the pump group velocity).  The coupling coefficient $\kappa$ has the form
\begin{equation}\label{eq:kappa}
\kappa = \epsilon_0 d_\mathrm{eff} \Theta \left( \frac{2 \hbar \omega_1 \omega_2 \omega_p Z_0^3}{n_1 n_2 n_p} \right)^{1/2},
\end{equation}
where $d_\mathrm{eff}$ is the effective nonlinear coefficient of the medium, $\Theta$ is the mode-overlap integral, and $Z_0 = \sqrt{\mu_0/\epsilon_0}$ is the impedance of free space \cite{pelc_long-wavelength-pumped_2011}.  The phase mismatch $\Delta k = k_2 - k_1 - k_p - 2\pi/\Lambda_G$, where $\Lambda_G$ is the poling period of the medium, and $\delta \nu_{jp} = u_j^{-1} - u_p^{-1}$.  We consider a pump laser with center frequency $\omega_p$ and spectral bandwidth $B$, consisting of $N_T = B/\Delta\nu$ oscillating modes, where $\Delta\nu$ is the free spectral range of the laser resonator.  Each mode has an amplitude $|c_n|$ and phase $\phi_n$, and, factoring out the carrier wave we can write the pump field as
\begin{equation}\label{eq:a_p}
a_p(t) = \sum_{n = -M}^{M} |c_n| e^{i(2\pi n\,\Delta\nu\,t + \phi_n)},
\end{equation}
and the instantaneous pump power is found as $P_p(t) = \hbar \omega_p |a_p(t)|^2$.

We now consider the case in which the pump laser bandwidth is smaller than the interaction bandwidth determined by $\Delta \omega = L/\delta \nu_{j,p}$.  Conceptually, we view this scenario in the time domain by considering that any temporal fluctuations of the pump occur on timescales longer than the interaction time defined by the group-velocity walkoff between the interacting waves.  Under these conditions, we can neglect the time-derivative terms in Eqns.~(\ref{eq:svea}):
\begin{align} \label{eq:nogvm}
\der{a_1}{z} &= -i \kappa a_p^\ast(t) a_2 e^{-i \Delta k z},\nonumber\\
\der{a_2}{z} &= -i \kappa a_p(t) a_1 e^{i \Delta k z}.
\end{align}
For a phasematched interaction with $\Delta k = 0$, these equations are readily solved to yield an input-output relation:
\begin{align} \label{eq:io}
a_1(L) &= a_1(0) \cos \kappa |a_p(t)| L + a_2(0) \sin \kappa |a_p(t)| L\nonumber\\
a_2(L) &= -a_1(0) \sin \kappa |a_p(t)| L + a_2(0) \cos \kappa |a_p(t)| L.
\end{align}
We consider the situation in which the converted signal at $\omega_2$ will be measured using a slow detector with bandwidth substantially lower than $B$.  This case naturally applies to the detection of single-photon-level signals, which are typically integrated for a timescale of at least several milliseconds using a counter. We can therefore write the time-averaged conversion efficiency as
\begin{equation}\label{eq:conv-eff}
\overline{\eta} = \frac{\overline{|a_2(L)|^2}}{\overline{|a_1(0)|^2}} = \overline{\sin^2\left(\sqrt{\eta_\mathrm{nor} P_p(t)} L\right) }
\end{equation}
where the overbar denotes a time-average.

For a single-mode pump, $P_p(t) = P_p$ and the conversion efficiency is not a fluctuating function of time.  For $P_p = P_\mathrm{max} = (\pi/2L)^2/\eta_\mathrm{nor}$, one attains unity conversion efficiency from $a_1$ to $a_2$.  The simplest multimode case is that of a two-mode laser, illustrated schematically in Fig.~\ref{fig:theory}(a).  The modes are separated by the laser-cavity free-spectral range $\Delta \nu$ and have equal magnitudes.  As shown in Fig.~\ref{fig:theory}(b), the instantaneous power $P_p(t)$ for this configuration is a sinusoidally oscillating function of time, with a frequency equal to the free spectral range $\Delta \nu$.  Because the instantaneous pump power is fluctuating rapidly on the time scale of the detector bandwidth, one will not observe unity conversion efficiency at an average pump power $\overline{P_p(t)} = P_\mathrm{max}$.  Rather, in the case of the two-mode pump laser, the time-averaged conversion efficiency peaks at 79\%.

In the experiment described below, we used a pump laser with approximately 300 longitudinal modes, with an approximately Gaussian envelope, illustrated in Fig.~\ref{fig:theory}(d).  Assuming random fixed phases between the oscillating modes, we calculated the instantaneous power and time-averaged conversion efficiency as was done in the two-mode case.  As can be seen in Fig.~\ref{fig:theory}(e), $P_p(t)$ is very spiky, and in Fig.~\ref{fig:theory}(f), we see that using this configuration one is only able to reach approximately 64\% conversion.

Our numerical results can be generalized using the tools of statistical optics.  A laser with $N_T > 5$ modes has intensity statistics that are nearly identical with that of thermal radiation \cite{goodman_statistical_1985}.  The probability distribution for the instantaneous power in a waveguide mode, $p_P(P; \bar{P})$, of a thermal source, is given by an exponential distribution
\begin{equation}\label{eq:thermal-p}
p_P(P; \bar{P}) = \left\{ \begin{array}{ll}
\bar{P}^{-1} \exp \left( -P/\bar{P} \right) & P \geq 0 \\
0 & \text{otherwise},
\end{array}
\right.
\end{equation}
where $\bar{P}$ is the average power.  From this distribution, we can calculate the expected conversion efficiency as an ensemble average
\begin{equation}\label{eq:eta-integral}
\braket{\eta(\bar{P})} = \int_0^\infty p_P(P; \bar{P}) \sin^2 \left( \sqrt{\eta_\mathrm{nor} P} L \right) \, dP.
\end{equation}
Evaluating Eq.~(\ref{eq:eta-integral}) for maximum $\braket{\eta(\bar{P})}$ numerically gives a result of 64\%, matching to within 1\% the result computed with Eq.~(\ref{eq:conv-eff}).

The theoretical results presented here differ from an earlier published description \cite{pan_single-photon_2008} in important ways that are described in detail in the Appendix.

\section{Experiment and Results}

We sought to use a multi-longitudinal-mode as a pump for low-noise upconversion of a telecom-band single-photon-level signal at 1.55~$\mu$m.  Due to a long tail in the Raman spectrum of LiNbO$_3$, pump wavelengths substantially longer than 1.8$\mu$m are needed to avoid noise photon generation by spontaneous anti-Stokes Raman scattering \cite{pelc_long-wavelength-pumped_2011}.

\begin{figure}
\includegraphics{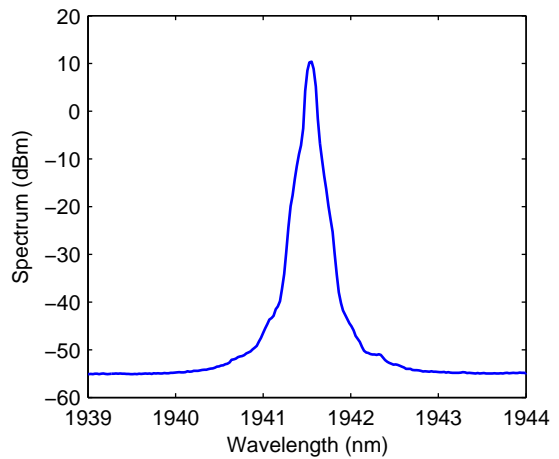}%
\caption{\label{fig:osa}Optical spectrum of Tm-doped fiber laser used in this experiment, after 20 dB optical attenuation.}
\end{figure}

We used a Tm-doped fiber laser (TmDFL) manufactured by AdValue Photonics (Tucson, AZ) as the pump source for this experiment.  We measured the laser spectrum using an optical spectrum analyzer; the results are shown in Fig.~\ref{fig:osa}.  The laser emits an average power of approximately 1 W into a spectral bandwidth of approximately 0.1~nm around a central wavelength of 1941.8~nm. We measured the free spectral range of the laser to be $\Delta \nu = 25.9$ MHz using an extended InGaAs photodiode and an rf spectrum analyzer.  We therefore estimate that approximately 300 modes of the cavity are participating in laser oscillation.

We fabricated PPLN waveguides via the reverse proton exchange technique \cite{parameswaran_highly_2002}.  The waveguides are 52 mm long and are poled with a QPM period of 19.6~$\mu$m.  The waveguides incorporated mode filters designed to match the mode size of SMF-28 optical fiber and were fiber-pigtailed with coupling losses of approximately 0.7 dB, and had a total fiber-to-output-facet throughput of -1.5 dB.  The waveguides were antireflection coated to avoid interference fringes and improve the system throughput.

We combined cw single-frequency radiation from a tunable telecom-band laser at $\lambda_1 = 1.54$~$\mu$m with the pump light at $\lambda_p = 1.94$~$\mu$m in a fiber-optic WDM and coupled it into the PPLN waveguide through the fiber pigtail.  The output depleted signal radiation at $\lambda_1$ and generated target radiation at $\lambda_2$ were collected with an aspheric lens ($f$ = 8~mm).  We measured the depletion of the input signal as a function of the pump power by coupling the light exiting the waveguide into an OSA and comparing the observed signal levels when the pump is turned on versus off.  This provides a calibration-free way of measuring the effects of the multimode pump laser on the conversion process.

\begin{figure}
\includegraphics{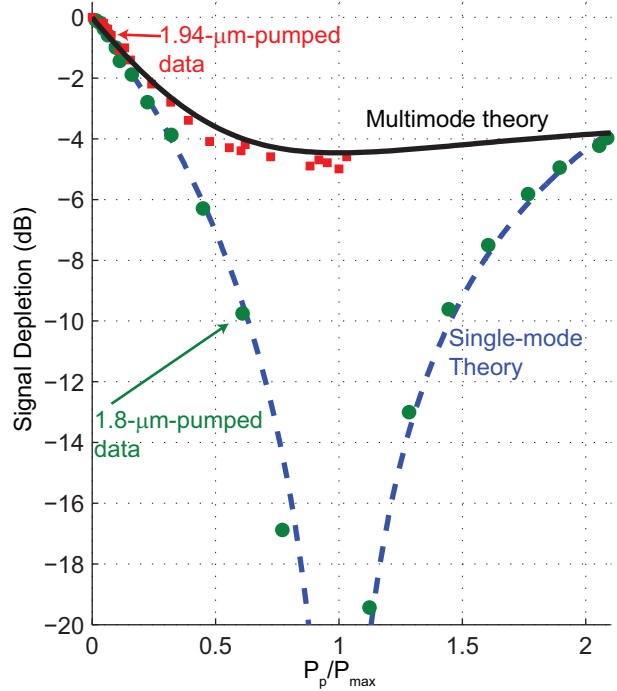}%
\caption{\label{fig:depletion} (Color on line) Signal depletion versus normalized pump power $P_p/P_\mathrm{max}$ for either single-mode optical pump at 1.48~$\mu$m (data from \cite{pelc_long-wavelength-pumped_2011}) or Tm:fiber laser pump at 1.94~$\mu$m.}
\end{figure}

Our measurement of the signal depletion, in logarithmic units, is shown in Fig.~\ref{fig:depletion} for both the case of the multimode Tm:fiber laser and a single-frequency 1.8-$\mu$m source described in \cite{pelc_long-wavelength-pumped_2011}.  In \cite{pelc_long-wavelength-pumped_2011}, a depletion level of 41 dB was observed, and matched a theoretical prediction of Eq.~(\ref{eq:conv-eff}) with high accuracy.  For the multimode-pumped case, we observed a maximum depletion of 5.0 dB, i.e. 70\% conversion efficiency.  The multimode theoretical curve in Fig.~\ref{fig:depletion} is calculated using the measured specifications of the TmDFL used in the experiment.  We calculated the conversion efficiency $\overline{\eta(P_p(t))}$ by taking an ensemble average over different values of random phases between the oscillating modes of the laser.  As can be seen, the conversion efficiency matches the theoretical prediction quite well with no free parameter other than the value of $P_\mathrm{max}$.

\begin{figure*}[htb]
%[width=0.98\textwidth]
\includegraphics{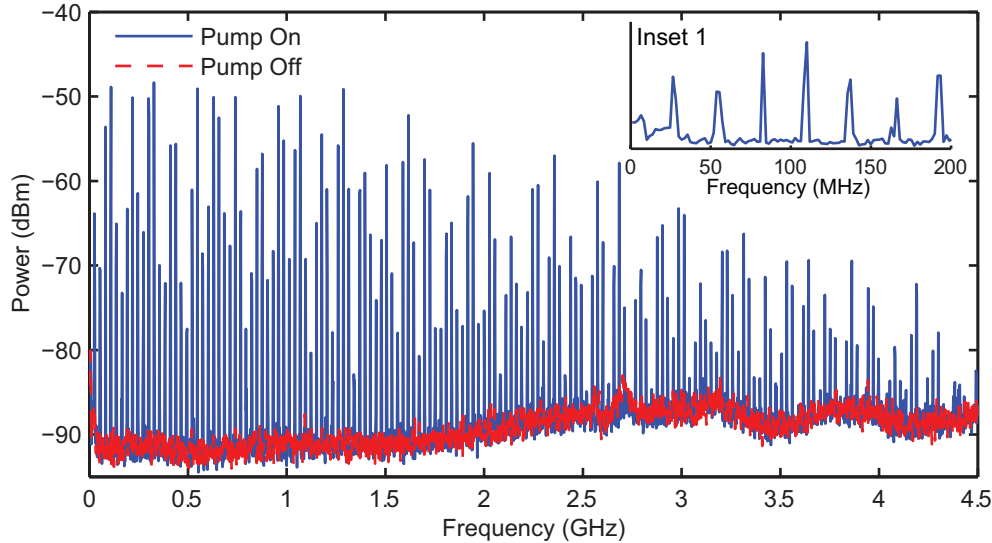}%
\caption{\label{fig:rf} (Color on line) RF spectrum of signal light at 1555.3 nm transmitted through the waveguide for 0 (pump off) or 3.47 dB (pump on) depletion.  When the pump is on, its rf spectral content is transferred onto the depleted signal, and also onto the generated target radiation.  Inset 1: zoom of rf spectrum between DC and 200 MHz, showing 7 rf spectral peaks with a mode spacing of 26 MHz.}
\end{figure*}

If one uses a multi-longitudinal-mode laser as the pump source in an upconverter, one expects the spectral structure of the pump to be transferred onto both the depleted input signal and the generated target radiation.  To confirm this, we observed the signal at $\lambda_1$ transmitted through the waveguide and compared the rf spectrum when the pump was turned on or off.  The signal was measured using an InGaAs photodiode with 10~GHz rf bandwidth, and coupled into an rf spectrum analyzer.  Our results are shown in Fig.~\ref{fig:rf}.  When the pump laser was turned off, the observed rf spectrum of the transmitted signal (dotted curve) showed some low frequency noise but then fell below the nearly white noise spectrum of the rf spectrum analyzer.  However, when the multimode pump was turned on, a comb of rf modes at integer multiples of $\Delta \nu$ appeared in the rf spectrum.  The inset contains the first seven rf spectral peaks, showing the $\Delta \nu \approx 26$~MHz free-spectral range of our pump laser.

\section{Discussion}

We demonstrated theoretically and experimentally that the instantaneous power fluctuations of a multi-mode pump limit the upconversion efficiency.  Using a pump with a spectral bandwidth of 0.1~nm at 1940~nm, containing roughly 300 oscillating modes, the time-averaged signal depletion saturated at approximately 5.0~dB, in good agreement with a semiclassical description of the upconversion process including interference effects due to the many pump modes beating against one another.

While we have experimentally focused on upconversion of cw single-frequency signals, the use of multi-longitudinal-mode pump lasers would also have deleterious effects when upconverting pulsed or modulated signals as would occur in a quantum communication system.  For phase-encoded signals (e.g. differential phase-shift quantum key distribution), the multimode pump would result in an upconverted signal in which the phase of sequential pulses would not be well controlled, likely corrupting the phase difference between pulses and destroying the qubit.

We note that the effects described with respect to upconversion would also apply to attempts to downconvert a quantum signal, to, for instance, convert single photons from trapped atoms or other quantum emitters in the visible or near-visible to the telecom band \cite{takesue_single-photon_2010, ikuta_wide-band_2011}.

A corollary of our findings is that optical frequency conversion can be an effective technique for measuring the spectral or temporal structure of lasers in frequency ranges where fast detectors are not well developed.  By mixing the multimode 1940-nm laser with C-band light in a nonlinear device, we were able to measure the spectral characteristics of the 1940-nm laser when no direct detector with 10-GHz bandwidth was available in that spectral region.  We anticipate that this technique may have applications in the characterization of emerging mid-IR laser sources.

\section*{Acknowledgements}

This work has been supported by the NNSF of China, the CAS, the National
Fundamental Research Program (under Grant No. 2011CB921300) and the National High Technology Research and Development Program (863 Program) of China under Grant No. 2009AA01A349. JSP and MMF acknowledge the United States AFOSR for their support under Grant FA9550-09-1-0233.

\section*{Appendix}

Upconversion of quantum states of light has been previously considered in a publication by Pan and coworkers \cite{pan_single-photon_2008}.  For a multimode pump, the authors give the Hamiltonian for the SFG process as
\begin{equation}\label{eq:pan-ham}
\hat{H} = i \hbar \sum_{i} \chi_i \left( \hat{a}_1 \hat{a}_2^\dagger - \mathrm{H.c.} \right)
\end{equation}
where the sum is over modes of the pump laser and the coupling constants
\begin{equation}\label{eq:pan-chi}
\chi_i = \chi' E_{pi},
\end{equation}
where $\chi'$ is determined by the $\chi^{(2)}$ of the medium and other experimental parameters, and the $\hat{a}_j$ are slowly-varying (i.e. optical carrier frequency has been factored out) photon annihilation operators.  The fact that the pump is not quantized as are the signal and SFG fields is justified by the fact that the pump field contains macroscopic numbers of photons.

We assert that the description implied by Eqs.~(\ref{eq:pan-ham})~and~(\ref{eq:pan-chi}) is incomplete, as it does not properly take into account the frequency difference between the pump modes.  To correctly take into account this fact, the coupling constants should be amended as
\begin{equation}\label{eq:chi-corr}
\chi_{i,\mathrm{corr}} = \chi' E_{pi} e^{i \Omega_i t}
\end{equation}
where $\Omega_i = \omega_{pi} - \omega_p$, where $\omega_{pi}$ is the optical frequency of pump mode $i$, and $\omega_p$ is the pump center frequency.  It is not possible, as the authors have done, to factor out each individual pump mode separately.  For a typical multimode laser $\omega_{p,i}$ and $\omega_{p,i-1}$ will differ by $2\pi \Delta \nu$.  Additionally, as the authors make no distinction between propagation and temporal evolution, the single-mode picture they use is necessarily incomplete to describe a scenario involving the propagation of a temporally fluctuating field.

An additional correction is needed following Eqns. (5) and (6) of \cite{pan_single-photon_2008}, where the coupling efficient $g$ is given.  Pan and coauthors give
\begin{equation}\label{eq:pan-Ep}
E_p^2 = \sum_i E_{pi}^2.
\end{equation}
A correct description of the square-modulus of the field would add the individual field modes before squaring:
\begin{equation}\label{eq:Ep-corr}
|E_p|^{2}_\mathrm{corr} = \left|\sum_i E_{pi} e^{i \Omega_i t}\right|^2
\end{equation}
In Pan and coauthors' description, while the individual pump modes are assumed to be at different frequencies, the fact that the total field intensity is computed incorrectly by an incoherent sum rather than a coherent sum wipes out the temporal fluctuations that occur as a result of the beating of the different pump longitudinal modes, as described by our Eq.~(\ref{eq:a_p}).

To properly describe the interaction of a multimode pump and a quantum signal via SFG, we assert that a model incorporating both spatial propagation and temporal effects is needed, as opposed to the single-variable treatment provided by \cite{pan_single-photon_2008}.  A proper multimode quantum optical picture of $\chi^{(2)}$ interactions is provided in \cite{loudon_quantum}.

%\bibliography{pelc_bib_all}
%\bibliography{}

\end{document}